\documentclass[aps,twocolumn,prl,amsmath]{revtex4}
\usepackage{epsfig}
\begin{document}

\noindent
{\bf Siano and Egger Reply:} \ 
In our recent Letter \cite{se}, we reported numerically exact
path-integral Monte Carlo (PIMC) simulation results for the
Josephson current through a  quantum dot in the Kondo regime.
Two of the central findings of Ref.~\cite{se} are as follows.
(1) Previous studies have
missed a minimum in the critical current as a function
of $\Delta/T_K$, where $\Delta_L=\Delta_R=\Delta$ is the BCS gap in the leads
and $T_K$ the Kondo temperature for normal leads, and (2)
the numerical renormalization group (NRG) method is 
inaccurate in the crossover regime ($\Delta/T_K\approx 1$).
These findings have been criticized by Choi, Lee, Kang and Belzig (CLKB)
\cite{comment}. First, they point out a 
mistake in our definition of $T_K$.  As discussed below,  this 
error in Ref.~\cite{se} does only affect numerical 
values for the various transition points
but neither the physical picture nor the conclusions drawn in Ref.~\cite{se}.
Their second point is that there are strong finite-$T$ effects.
Here one should first note that PIMC is inherently a finite-$T$ 
technique but has the benefit of exactness. 
While we did not claim true ground-state results in Ref.~\cite{se}, 
finite-$T$ predictions are relevant for experiments, where 
$T\approx 100$mK corresponding to our value for realistic applications.
Moreover, our finite-$T$ results reproduce the analytical $T=0$ solution for
$\Delta/T_K\gg 1$ \cite{se}, in contrast to the assertions of CLKB. 
Of course, at $T/\Delta=0.1$, significant
thermal smearing in the $I(\phi)$ curves is present,
but this is obvious and also seen in our PIMC simulations. 
Interestingly, a direct comparison of the finite-$T$ NRG
results shown in Fig.~1 of Ref.~\cite{comment} to exact PIMC simulation data 
reveals that NRG is not of sufficient accuracy for this
problem.  Problems of NRG for the case $\Delta_L=\Delta_R$ under discussion
here have been mentioned also in Ref.~\cite{oguri}.

\begin{figure}
\scalebox{0.25}{
\rotatebox{270}{
\includegraphics{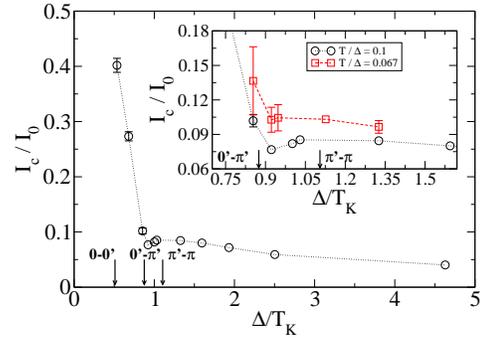}
}}
\caption{
\label{f1} 
Plot of Fig.~5 of Ref.~\cite{se} with corrected $\Delta/T_K$ units. In the
inset, new PIMC data at lower $T$ have been added.
}
\end{figure}

\begin{figure}
\scalebox{0.25}{
\rotatebox{0}{
\includegraphics{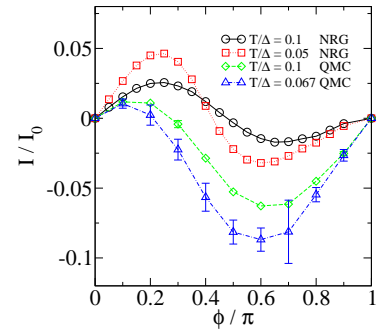}
}}
\caption{
\label{f2}  Direct comparison between finite-$T$ NRG and PIMC results at
$U/\Delta=14$ and $\Gamma/\Delta=2.82$. Vertical bars (or the symbol
size) indicate standard MC error estimates.
}
\end{figure}

As pointed out by CLKB,
by letting $\Gamma\to 2\Gamma$ in Eq.~(1) of Ref.~\cite{se} (and only there!),
the correct $T_K$ emerges.
This change implies that equations (8) to (11) in Ref.~\cite{se} now read:
\begin{eqnarray*}
(\Delta/T_K)_{00'} &=& 0.51\pm 0.01, \\
(\Delta/T_K)_{0'\pi'}&=& 0.875\pm 0.005, \\
(\Delta/T_K)_{\pi' \pi}&=& 1.105\pm 0.005, \\
(\Delta/T_K)_{\rm min}&=& 0.92\pm 0.03. 
\end{eqnarray*}
Furthermore, in Fig.~2 of Ref.~\cite{se}, we have $\Delta/T_K\approx 2.5$,
while in Fig.~3 the corresponding values were $\Delta/T_K=0.53, 0.68,$ and
$0.85$.
Likewise, in Fig.~4 the correct values should read
$\Delta/T_K=0.85,1.03, 1.13,$ and $11.6$. 
The corrected Fig.~5 of Ref.~\cite{se} is shown here in 
Fig.~\ref{f1}, where we also added new PIMC data  obtained at a 
lower temperature in the inset.
Fortunately,  these changes in the numerical values of the various transition
points do not affect our main conclusions.
In particular, Fig.~\ref{f1} shows that the 
minimum in $I_c$ as a function of $\Delta/T_K$ persists upon 
lowering temperature,
and is not necessarily wiped out by thermal fluctuations.  
In Fig.~\ref{f2}, we show an explicit comparison of exact PIMC results
 to the finite-$T$ NRG data
of CLKB, using precisely the same microscopic parameters as in 
 Ref.~\cite{comment}.  Figure \ref{f2} indicates
that the well-established high accuracy of NRG in treating ``normal'' 
Kondo problems is lost due to BCS gaps in the leads \cite{oguri}.
In the crossover regime, $\Delta/T_K\approx 1$, NRG is therefore unreliable.
We conclude that, apart from the mistake regarding
$T_K$ posing no fundamental problem,
the criticism of CLKB is unfounded.

\vspace{0.1cm}
\noindent
F. Siano and R. Egger \\
{\small Institut f\"ur Theoretische Physik, Heinrich-Heine-Universit\"at,
D-40225 D\"usseldorf, Germany }

\end{document}